\documentclass[english,10pt,aps,pra,twocolumn,groupedaddress,showkeys,floatfix]{revtex4-1}

\usepackage[utf8]{inputenc}
\usepackage{amsmath}
\usepackage{amssymb}
\usepackage{amsfonts}
\usepackage{bm}
\usepackage{bbm}
\usepackage{epsfig}
\usepackage{grffile}
\usepackage{graphics}
\usepackage{extarrows}
\usepackage{siunitx}

\usepackage[usenames,dvipsnames]{color}
\definecolor{dblue}{rgb}{0,0.1,.6}

\usepackage[colorlinks=true,citecolor=dblue,linkcolor=dblue,urlcolor=dblue]{hyperref}
\usepackage[all]{hypcap}

\newcommand{\ud}{\mathrm{d}}
\newcommand{\id}{\mathbbm{1}}
\newcommand{\Tr}{\operatorname{Tr}}
\newcommand{\bra}{\langle}
\newcommand{\ket}{\rangle}
\newcommand{\mc}[1]{\mathcal{#1}}
\newcommand{\pdag}{{\phantom{\dag}}}

\renewcommand{\H}{\mc{H}}
\newcommand{\hH}{\hat{H}}
\newcommand{\hN}{\hat{N}}
\newcommand{\hS}{\hat{S}}
\renewcommand{\vr}{\vec{r}}
\newcommand{\vk}{\vec{k}}
\newcommand{\vn}{\vec{n}}
\newcommand{\vz}{\vec{z}}
\newcommand{\hx}{\hat{x}}
\newcommand{\hp}{\hat{p}}
\newcommand{\ha}{\hat{a}}
\newcommand{\hb}{\hat{b}}
\newcommand{\tb}{\tilde{b}}
\newcommand{\tn}{\tilde{n}}
\newcommand{\dm}{{\hat{\varrho}}}
\renewcommand{\vec}[1]{{\boldsymbol{#1}}}
\newcommand{\A}{\mc{A}}
\newcommand{\B}{\mc{B}}
\newcommand{\N}{\mc{N}}

\newcommand{\veps}{\varepsilon}
\newcommand{\vphi}{\varphi}
\newcommand{\gs}{\text{gs}}
\newcommand{\therm}{\text{th}}
\newcommand{\nsing}{\text{ns}}
\newcommand{\sing} {\text{s}}
\newcommand{\sq} {\text{sq}}

\newcommand{\tmat}[1] {\begin{bmatrix}#1\end{bmatrix}}
\newcommand{\tvec}[1] {\begin{pmatrix}#1\end{pmatrix}}

\newcommand{\duke} {Department of Physics, Duke University, Durham, North Carolina 27708, USA}

\newcommand{\Title} {Scaling functions for eigenstate entanglement crossovers in harmonic lattices}
\newcommand{\Authors}
{
\author{Thomas Barthel}
\affiliation{\duke}
\author{Qiang Miao}
\affiliation{\duke}
}
\newcommand{\Date} {July 27, 2021}

\begin{document}

\title{\Title}
\Authors

\begin{abstract}
For quantum matter, eigenstate entanglement entropies obey an area law or log-area law at low energies and small subsystem sizes and cross over to volume laws for high energies and large subsystems. This transition is captured by crossover functions, which assume a universal scaling form in quantum critical regimes. We demonstrate this for the harmonic lattice model, which describes quantized lattice vibrations and is a regularization for free scalar field theories, modeling, e.g., spin-0 bosonic particles. In one dimension, the groundstate entanglement obeys a log-area law. For dimensions $d\geq 2$, it displays area laws, even at criticality. The distribution of excited-state entanglement entropies is found to be sharply peaked around subsystem entropies of corresponding thermodynamic ensembles in accordance with the eigenstate thermalization hypothesis. 
Numerically, we determine crossover scaling functions for the quantum critical regime of the model and do a large-deviation analysis. We show how infrared singularities of the system can be handled and how to access the thermodynamic limit using a perturbative trick for the covariance matrix. Eigenstates for quasi-free bosonic systems are not Gaussian. We resolve this problem by considering appropriate squeezed states instead. For these, entanglement entropies can be evaluated efficiently.
\end{abstract}

\date{\Date}
\maketitle

\section{Introduction}\label{sec:intro}
Consider a bipartite system in a pure state $|\psi\ket\in\H_\A\otimes\H_\B$. Then, quantum non-locality can be quantified by the von Neumann entanglement entropy 
\begin{equation}\label{eq:vNentang}
	S=-\Tr\dm_\A\ln\dm_\A,
\end{equation}
where $\dm_\A=\Tr_\B|\psi\ket\bra\psi|$ is the reduced density matrix of subsystem $\A$  \cite{Schumacher1995-51,Jozsa1994-41,Bennett1996,Nielsen2000}. Specifically, we consider compact subsystems $\A$ with linear size $\ell$, i.e., volume $\ell^d$ where $d$ is the number of spatial dimensions, and we assume $\A$ to be much smaller than the rest $\B$. For typical condensed matter systems, the groundstate entanglement entropy obeys an area law, $S\sim \ell^{d-1}$ \cite{Callan1994-333,Latorre2004,Calabrese2004,Plenio2005,Cramer2006-73,Hastings2007-76,Brandao2013-9,Cho2018-8,Kuwahara2020-11}, or a log-area law, $S\sim \ell^{d-1}\log\ell$ \cite{Wolf2005,Gioev2005,Barthel2006-74,Li2006,Lai2013-111}. According to quantum typicality \cite{Popescu2006-2,Goldstein2006-96,Gemmer2004}, generic states should, however, obey a volume law, i.e., have an extensive entanglement entropy $S\sim \ell^{d}$. As discussed in our recent contribution \cite{Miao2019_05}, entanglement entropies of energy eigenstates, in particular, cross over from the groundstate scaling at low energies and small subsystem sizes to an extensive scaling at high energies and large subsystem sizes. Previous work considered states with few-particle excitations, i.e., vanishing excitation-energy density, and found that, up to subleading corrections, they still obey (log-)area laws \cite{Das2006-73,Das2008-77,Masanes2009-80,Alcaraz2011-106,Berganza2012-01,Moelter2014-10,Moudgalya2018-98}. The same scaling was found for special classes of (rare) highly excited states \cite{Das2006-73,Alba2009-10,Moudgalya2018-98,Vafek2017-3} which, in some cases, can be interpreted as ground states of other local Hamiltonians. Extensive entanglement entropies have been demonstrated for broad classes of highly excited states in Refs.~\cite{Alba2009-10,Ares2014-47,Storms2014-89,Keating2015-338}.
Extensive scaling of the \emph{average} eigenstate entanglement was shown for quasi-free fermionic systems and chaotic Hamiltonians \cite{Vdimar2017-119,Vidmar2017-119b,Vidmar2018-121,Lu2019-99,Huang2019-938}.
\begin{figure}[t]
	\includegraphics[width=0.9\columnwidth]{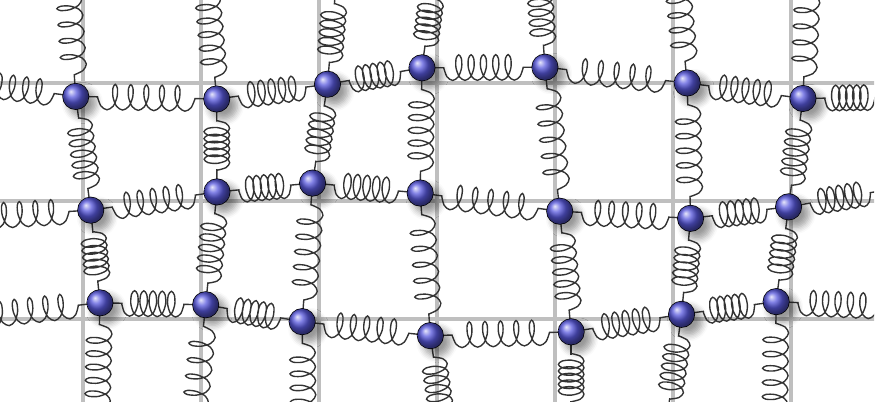}
	\caption{\label{fig:HLM}\textbf{Mechanical analog of the harmonic lattice model.} Oscillators of eigenfrequency $\omega$ on a $d$-dimensional lattice are coupled by springs of strength $\kappa$. The Hamiltonian has $d$ uncoupled terms $\hH=\hH_x+\hH_y\dots$, each corresponding to the harmonic lattice model in Eq.~\eqref{eq:HLM-r}.}
\end{figure}

In the recent contribution \cite{Miao2019_05}, we pointed out that the crossover from groundstate scaling to volume laws can be deduced using the eigenstate thermalization hypothesis (ETH). The entanglement entropies of (almost) all eigenstates converge, in the thermodynamic limit, to the subsystem entropy of global thermal ensembles with the same energy density. They are hence captured by a single crossover function, and, in the quantum critical regime \cite{Sachdev2011}, these crossover functions become universal scaling functions. In Ref.~\cite{Miao2019_05}, we used results of conformal field theory (CFT) to obtain analytical expressions for the crossover scaling functions of critical one-dimensional (1d) systems and of Fermi liquids in $d\geq 2$ dimensions. Furthermore, the crossover function for a gapped 1d system (massive Dirac fermions) was determined numerically. The applicability of weak ETH for the subsystem entropies was confirmed by numerics for large non-interacting fermionic systems. In Ref.~\cite{Miao2020_10}, the applicability of ETH and the CFT scaling function were confirmed for integrable and non-integrable spin chains.

In this paper, we consider translation-invariant harmonic lattice models
\begin{equation}\label{eq:HLM-r}
	\hH=\frac{1}{2}\sum_{\vr}(\hp_\vr^2+\omega^2\hx_\vr^2)+\frac{1}{2}\,\kappa\sum_{\bra \vr,\vr'\ket}(\hx_\vr-\hx_{\vr'})^2
\end{equation}
with $[\hx_\vr,\hp_{\vr'}]=i\delta_{\vr,\vr'}$, and determine scaling functions for the eigenstate entanglement crossover from the groundstate scaling to volume laws.
The models describe, e.g., quantized lattice vibrations in solids (phonons), which follow Bose-Einstein statistics. The continuum limit is the free scalar (Klein-Gordon) quantum field theory 
\begin{equation}\label{eq:H-KG}
	\hH=\frac{1}{2}\int\ud^d r\Big[ \hat{\pi}^2+\omega^2\hat{\phi}^2+\kappa\big(\partial_\vr \hat{\phi}\big)^2\Big]
\end{equation}
with $[\hat{\phi}(\vr),\hat{\pi}(\vr')]=i\delta(\vr-\vr')$. In particle physics with $d=3$ and $\kappa=1$, it is used to describe spin-0 bosons of mass $\omega$. In the gapped regime $\omega>0$, the groundstate entanglement entropy obeys the area law.
At the quantum critical point $\omega=0$, the system has gapless excitations and the groundstate entanglement entropy follows the log law $S_\gs\sim \frac{c}{3}\ln \ell$ for $d=1$ \cite{Srednicki1993,Callan1994-333,Holzhey1994-424,Calabrese2004}, but still follows an area law $S_\gs\sim\alpha\,\ell^{d-1}$ for $d>1$ \cite{Srednicki1993,Callan1994-333,Barthel2006-74,Casini2009-42}. See Refs.~\cite{Eisert2008,Latorre2009,Laflorencie2016-646} for reviews on groundstate entanglement scaling.

The ETH \cite{Deutsch1991-43,Srednicki1994-50,Rigol2008-452,Biroli2010-105,Beugeling2014-89,Kim2014-90,Alba2015-91,Lai2015-91,Dymarsky2018-97,Deutsch2018-81} implies that local expectation values of all (strong ETH) or at least the majority of all (weak ETH) energy eigenstates are indistinguishable from expectation values of corresponding microcanonical ensembles with the same energy. Deviations vanish in the thermodynamic limit. In contrast to strong ETH, weak ETH allows for an exponentially small number of untypical eigenstates \cite{Yoshizawa2018-120}. While strong ETH is difficult to establish in a general way, weak ETH \cite{Biroli2010-105,Mori2016_09,Iyoda2017-119} applies very generally and can be explained rather easily, either through an argument on the spatial decay of correlations or a resolution limitation effect in momentum space. See Ref.~\cite{Miao2019_05}.
We show and use that, due to the ETH, the entanglement entropies of excited states can be deduced from subsystem entropies of corresponding thermodynamic ensembles. Exploiting the equivalence of thermodynamic ensembles for large systems \cite{Lebowitz1967-153,Ruelle1969,Mueller2015-340,Tasaki2018-172}, we employ the grand-canonical ensemble (GCE) $\dm=\exp(-\beta\hH)/Z$ to deduce entanglement entropies in excited states with energy $E(\beta)=\bra\hH\ket_\beta$ and phonon number $N(\beta)=\bra\hN\ket_\beta$. For simplicity, we keep the chemical potential at $\mu=0$. Entanglement entropies of typical eigenstates are very close to GCE subsystem entropies. Hence, they are extensive for large subsystems,
\begin{equation}\label{eq:crossover}
	S(\ell,E) \stackrel{\text{typical}}{\approx}
	S_\text{GCE}(\ell,\beta) \xrightarrow{\ell\gg\xi }\ell^d\, s_\therm(\beta),
\end{equation}
where $s_\therm(\beta)$ denotes the thermodynamic entropy density, and $\xi$ is the thermal correlation length. More specifically, $S$ crosses over from the groundstate scaling at small $\ell$  to extensive scaling at large $\ell$. The crossover length increases with decreasing energy ($\beta^{-1}$). That the groundstate scaling is recovered at small $\ell$ is due to the momentum-space resolution limitation \cite{Miao2019_05}: Limiting measurements to a small subsystem corresponds to coarse-graining in momentum space on a scale $\sim 1/\ell$, due to which excited states become indistinguishable from the ground state.

The paper is organized as follows. In Sec.~\ref{sec:Diag}, the harmonic lattice model is diagonalized, and we give the covariance matrices that characterize thermal equilibrium states. Section~\ref{sec:Entropy} shows how entanglement entropies and finite-temperature subsystem entropies can be computed from covariance matrices. The critical 1d and 2d harmonic lattice models feature ultraviolet and, more importantly, infrared singularities. We discuss in Sec.~\ref{sec:Singular} how these can be regularized and how we can access the thermodynamic limit while retaining scale invariance. Finally, scaling functions for the eigenstate entanglement crossovers are determined numerically in Sec.~\ref{sec:Crossovers}. Eigenstates for quasi-free bosonic systems are not Gaussian, and hence it is usually computationally expensive to assess their entanglement entropies. We resolve this problem by considering appropriate squeezed states instead. They are discussed in Sec.~\ref{sec:Squeezed} and used in Sec.~\ref{sec:assertETH} to assert the applicability of the ETH for the considered systems, including a large-deviation analysis. We conclude in Sec.~\ref{sec:Conclusion}, also commenting on the universality of the obtained scaling functions.

\section{Diagonalization of the harmonic lattice model}\label{sec:Diag}
The position and momentum operators in the harmonic lattice model \eqref{eq:HLM-r} obey the canonical commutation relations
\begin{equation*}
	[\hx_\vr,\hp_{\vr'}]=i\delta_{\vr,\vr'},\quad
	[\hx_\vr,\hx_{\vr'}]=0,\quad
	[\hp_\vr,\hp_{\vr'}]=0.
\end{equation*}
The translation invariance can be utilized to transform to a system of uncoupled harmonic oscillators
\begin{equation}\label{eq:HLM-Hk}
	\hH=\frac{1}{2}\sum_{\vk}(\hp_\vk^\dag\hp^\pdag_\vk+\veps_\vk\hx_\vk^\dag\hx^\pdag_\vk),
\end{equation}
where $\veps_\vk$ is specified below and
\begin{subequations}\label{eq:xk-pk}
\begin{align}
	\hx_\vk &= \frac{1}{\sqrt{\N}}\sum_\vr e^{-i\vk\cdot\vr}\hx_\vr= \hx_{-\vk}^\dag,\\
	\hp_\vk &= \frac{1}{\sqrt{\N}}\sum_\vr e^{-i\vk\cdot\vr}\hp_\vr= \hp_{-\vk}^\dag
\end{align}
\end{subequations}
such that 
\begin{equation}\label{eq:CCR-k}
	[\hx_\vk^\dag,\hp^\pdag_{\vk'}]=i\delta_{\vk,\vk'},\quad
	[\hx_\vk^\dag,\hx^\pdag_{\vk'}]=0,\quad
	[\hp_\vk^\dag,\hp^\pdag_{\vk'}]=0.
\end{equation}
The final step of the diagonalization is to define bosonic ladder operators
\begin{equation}\label{eq:bk}
	\hb_\vk=\frac{1}{\sqrt{2}}\left(\veps_\vk^{1/2}\hx_\vk+i\veps_\vk^{-1/2}\hp_\vk\right)
\end{equation}
such that
\begin{equation}\label{eq:HLM-diag}
	\hH=\sum_{\vk}\veps_\vk\left(\hb_\vk^\dag\hb^\pdag_\vk+1/2\right).
\end{equation}

For a $d$-dimensional system with lattice basis vectors $\vec{a}_i$, the phonon dispersion relation is
\begin{equation}
	\veps_\vk=\sqrt{\omega^2+4\kappa\sum_{i=1}^d\sin^2\left(\frac{\vk\cdot\vec{a}_i}{2}\right)}.
\end{equation}
For clarity, we will assume a square lattice on a $d$-dimensional torus in the following, i.e., $[\vec{a}_i]_j=\delta_{i,j}$ and $k_i=0,\frac{2\pi}{L_i},\dotsc,(L_i-1)\frac{2\pi}{L_i}$, $i=1,\dotsc,d$, where $\{L_i\}$ are the circumferences of the torus. The total number of sites is hence $\N=\prod_{i=1}^d L_i$.
In the low-energy regime, relative displacements of neighbors are small, and one can take the continuum limit of the model, yielding the free scalar (Klein-Gordon) quantum field theory \eqref{eq:H-KG}.

The energy gap of the system vanishes for $\omega\to 0$, closing at $\vk=\vec{0}$. For numerical computations, one can avoid problems with the zero-momentum mode by switching to antiperiodic boundary conditions, implying a shift of the momenta $k_i$ by ${\pi}/{L_i}$. Antiperiodic boundary conditions correspond to a coupling like $(\hx_1+\hx_L)^2$ across the boundary. For $\omega=0$, this keeps the oscillators from flying off to $x_\vr\to\pm\infty$ and imposes a finite-size gap $\propto 1/L$. Further issues with infrared singularities of the model occurring in 1d and 2d are discussed in Sec.~\ref{sec:Singular}. 

The fact that the model is quadratic in $\hx_\vr$ and $\hp_\vr$ implies that equilibrium states $\dm=\frac{1}{Z}\,e^{-\beta\hH}$ are Gaussian states and, according to Wick's theorem \cite{Fetter1971}, all observables can hence be computed from the the single-particle Green's functions. In particular, we will employ the covariance matrices
\begin{subequations}\label{eq:CovMatrices}
\begin{align}
	G^x_{\vr,\vr'}&:=\bra\hx_\vr\hx_{\vr'}\ket_\beta
	 = \frac{1}{\N}\sum_{\vk,\vk'}e^{-i\vk\cdot\vr}e^{i\vk'\cdot\vr'}\bra\hx_\vk^\dag\hx_{\vk'}^\pdag\ket_\beta\nonumber\\
	 &= \frac{1}{2\N}\sum_\vk \cos(\vk\cdot\Delta\vr)\,\frac{1}{\veps_\vk}\,\coth\left(\frac{\beta\veps_\vk}{2}\right),\\
	G^p_{\vr,\vr'}&:=\bra\hp_\vr\hp_{\vr'}\ket_\beta
	 = \frac{1}{\N}\sum_{\vk,\vk'}e^{-i\vk\cdot\vr}e^{i\vk'\cdot\vr'}\bra\hp_\vk^\dag\hp_{\vk'}^\pdag\ket_\beta\nonumber\\
	 &= \frac{1}{2\N}\sum_\vk \cos(\vk\cdot\Delta\vr)\,\veps_\vk\,\coth\left(\frac{\beta\veps_\vk}{2}\right),\\
	G^{xp}_{\vr,\vr'}&:=\bra\hx_\vr\hp_{\vr'}\ket_\beta = \frac{i}{2}\delta_{\vr,\vr'}.
\end{align}
\end{subequations}
These are functions of $\Delta\vr=\vr-\vr'$ only, due to the translation invariance.

\section{Subsystem density matrices and entropies}\label{sec:Entropy}
Let $\dm_\A=\Tr_\B\dm$ denote the reduced density matrix for subsystem $\A$, when the total system is in a Gaussian state $\dm$. As pointed out in Sec.~\ref{sec:Diag}, the Gaussian states $\dm$ with vanishing first moments $\bra\hx_\vr\ket$ and $\bra\hp_\vr\ket$ are fully characterized by the covariance matrices $G^x_{\vr,\vr'}=\bra\hx_\vr\hx_{\vr'}\ket$, $G^p_{\vr,\vr'}=\bra\hp_\vr\hp_{\vr'}\ket$, and $G^{xp}_{\vr,\vr'}=\bra\hx_\vr\hp_{\vr'}\ket$. Expectation values of arbitrary observables can be computed from $G$ through Wick's theorem \cite{Fetter1971}. It follows further, that $\dm_\A$ is a Gaussian state, characterized by the subsystem covariance matrices
\begin{equation}
	g^x:=G^x|_\A,\quad
	g^p:=G^p|_\A,\quad
	g^{xp}:=G^{xp}|_\A,\quad
\end{equation}
i.e., the restriction of the covariance matrices to sites $\vr,\vr'$ in subsystem $\A$. For a subsystem with $\N_\A$ sites, these are $\N_\A\times \N_\A$ matrices.

Let us now employ a canonical transformation $T$ for the position and momentum operators in the subsystem to diagonalize $g$ in the sense that
\begin{equation*}
	\begin{bmatrix}g^x&g^{xp}-\frac{i}{2}\\(g^{xp}-\frac{i}{2})^\dag&g^p\end{bmatrix}
	=T^\top \begin{bmatrix}\nu&\\&\nu\end{bmatrix} T,
	\quad  T\in\operatorname{Sp}(\N_\A),
\end{equation*}
For a canonical transformation, $T$ has to be a symplectic matrix. The diagonal matrix $\nu$ contains the symplectic eigenvalues $\nu_k\geq 1/2$. Such a transformation always exists according to the Williamson theorem \cite{Williamson1936-58,Simon1999-40}. For cases where $g^{xp}=\frac{i}{2}\id$, as for the equilibrium states $\dm=\frac{1}{Z}\,e^{-\beta\hH}$ in Sec.~\ref{sec:Diag}, the diagonalization can be simplified:
\begin{gather}\label{eq:gammaDiag-xp}
	M:=\sqrt{g^p}\,g^x\sqrt{g^p}\,
	\stackrel{\operatorname{diag}}{=}\, O^\top \nu^2\, O,\quad
	T=T_x\oplus T_p\\\nonumber
	\text{with} \quad
	T_x=\nu^{\frac{1}{2}}\,O\,(g^p)^{-\frac{1}{2}} \ \,\text{and}\, \
	T_p=\nu^{-\frac{1}{2}}\,O\,(g^p)^{\frac{1}{2}}.
\end{gather}

The transformation $T$ yields new position and momentum operators and corresponding bosonic ladder operators $\tb_k^\dag$, $\tb_k$  that create and destroy particles in $\A$ such that 
\begin{equation}\label{eq:CovDiag}
	\bra\tb^\dag_k\tb^\pdag_{k'}\ket=\delta_{k,k'}\,(\nu_k-1/2)\quad\text{and}\quad 
	\bra\tb_k\tb_{k'}\ket=0.
\end{equation}
As a Gaussian state with vanishing first moments, $\dm_\A$ is the exponential of a quadratic form in the ladder operators. According to Eq.~\eqref{eq:CovDiag}, the subsystem density matrix hence takes the form
\begin{equation*}
	\dm_\A=\prod_k \dm_k\quad\text{with}\quad
	\dm_k=\frac{1}{\nu_k+{1}/{2}}\left(\frac{\nu_k-{1}/{2}}{\nu_k+{1}/{2}}\right)^{\tb_k^\dag\tb_k^\pdag}.
\end{equation*}
Finally, we obtain the subsystem entropy in terms of the symplectic eigenvalues
\begin{gather}
	S_\A=-\Tr\dm_\A\ln\dm_\A = -\sum_k \Tr\dm_k\ln\dm_k = \sum_k h(\nu_k),\nonumber\\ \label{eq:Snu}
	h(\nu)\textstyle =\left(\nu+\frac{1}{2}\right)\ln\left(\nu+\frac{1}{2}\right)-\left(\nu-\frac{1}{2}\right)\ln\left(\nu-\frac{1}{2}\right).
\end{gather}
For pure states $\dm$, this is the von Neumann entanglement entropy.

\section{Singularities, regularization, and thermodynamic limit}\label{sec:Singular}
In the following, we will only concern ourselves with the the harmonic lattice model \eqref{eq:HLM-r} at the critical point $\omega=0$. Also, we can set $\kappa=1$ without loss of generality because it can be eliminated through a canonical transformation $\hx_\vr\mapsto\kappa^{-1/4}\,\hx_\vr$, $\hp_\vr\mapsto\kappa^{1/4}\,\hp_\vr$, and $\hH\mapsto\hH/\sqrt{\kappa}$. The low-energy field theory \eqref{eq:H-KG} is then rotation- and scale-invariant.

Ultraviolet and infrared singularities complicate the study of the (critical) harmonic lattice model. Here, we discuss how these can be handled and how we can access the thermodynamic limit while retaining scale invariance.

In the thermodynamic limit $L_i\to\infty$, the covariance matrices \eqref{eq:CovMatrices} take the form
\begin{subequations}\label{eq:CovMatrices-TD}
\begin{align} \label{eq:CovMatrices-TDa}
	G^x_{\Delta\vr}
	 &\textstyle= \frac{1}{2(2\pi)^d}\int\ud^d k\, \cos(\vk\cdot\Delta\vr)\frac{1}{\veps_\vk}\coth\left(\frac{\beta\veps_\vk}{2}\right),\\ \label{eq:CovMatrices-TDb}
	G^p_{\Delta\vr}
	 &\textstyle= \frac{1}{2(2\pi)^d}\int\ud^d k\, \cos(\vk\cdot\Delta\vr)\,\veps_\vk\coth\left(\frac{\beta\veps_\vk}{2}\right).
\end{align}
\end{subequations}

In the continuum limit \eqref{eq:H-KG}, the dispersion is linear for arbitrarily large momenta, i.e., $\veps_\vk=|\vk|$. The resulting divergent ultraviolet behavior is most pronounced in $G^p_{\Delta\vr=\vec{0}}\sim \int\ud k\, k^d$. This singularity can be resolved by imposing an ultraviolet cutoff $k_{\operatorname{max}}$, or by studying the model on a lattice. As described in Sec.~\ref{sec:Diag}, we follow the second approach and the integrals \eqref{eq:CovMatrices-TD} run over the Brillouin zone $[-\pi,\pi)^d$.

While the thermodynamic limit for $G^p$ is well-defined, $G^x$ has an infrared singularity for 1d systems at nonzero temperatures, 1d systems in the ground state, and 2d systems at nonzero temperatures. The $d$ dependence is due to the density of states; for small energies it is $g(\veps)\propto\veps^{d-1}$. See Appendix~\ref{sec:Sth}. As far as we know, the infrared singularities have not really been discussed in the literature so far; for 1d ground states, they have of course been noticed \cite{Srednicki1993,Callan1994-333,Skrovseth2005-72,Evenbly2010-12,Mallayya2014-90}.
At small momenta, the integrand in Eq.~\eqref{eq:CovMatrices-TDa} is $\sim {2}/({\beta k^2})$ for nonzero temperatures and it is $\sim {1}/{k}$ for the ground state. The singularities can be regularized by introducing an infrared cutoff $k_0$ or, equivalently, choosing a finite linear system size $L$ ($k_0\sim 2\pi/L$). Since we want to derive scaling functions for subsystem entropies at nonzero temperature, we also want the low-energy/long-distance features of the regularized $G^x$ to be scale invariant. This can be achieved by choosing a momentum cutoff that is proportional to temperature,
\begin{equation*}
	k_0=q_0/\beta,\quad q_0=const.
\end{equation*}
or, equivalently, finite system sizes $L\propto\beta$.
For low energies, where $\veps_\vk\approx |\vk|$, we then have indeed
\begin{equation*}
	G^x_{\Delta\vr}\to
	 \frac{\beta^{1-d}}{2(2\pi)^d}\int_{|\vec{q}|>q_0}\hspace{-2ex}\ud^d q\, \cos\left(\frac{\vec{q}\cdot\Delta\vr}{\beta}\right)\frac{1}{|\vec{q}|}\coth\left({|\vec{q}|}/{2}\right),
\end{equation*}
i.e., $\beta^{d-1}G^x$ becomes a function of $\Delta\vr/\beta$ only instead of $\Delta\vr$ \emph{and} $\beta$.

We can split $G^x_{\Delta\vr}$ into a non-singular ($q_0$ independent) part $G^{x(\nsing)}_{\Delta\vr}$ and a singular part $G^{x(\sing)}$, which, importantly, is independent of $\Delta\vr$. At nonzero temperatures, the singular part is
\begin{subequations}\label{eq:GxSing}
\begin{align}\label{eq:GxSing1d}
	G^{x(\sing)}&=\frac{1}{\pi q_0}
	&\text{for 1d},\\ \label{eq:GxSing2d}
	G^{x(\sing)}&=\frac{1}{2\pi\beta}\ln\left(\frac{1}{q_0}\right)\ 
	&\text{for 2d}.
\end{align}
\end{subequations}

Let us deduce the corresponding non-singular and singular contributions to subsystem entropies. For the $\N_\A\times \N_\A$ subsystem covariance matrices from Sec.~\ref{sec:Entropy}, we have
\begin{equation}\label{eq:gx-perturb}
	g^x=\frac{1}{\lambda}\vec{v}\vec{v}^\top+g^{x(\nsing)}\quad \text{with}\quad \frac{1}{\lambda}:=G^{x(\sing)},
\end{equation}
where $g^{x(\nsing)}$ is the restriction of $G^{x(\nsing)}$ to the subsystem $\A$ and $\vec{v}:=(1,1,\dotsc,1)^\top$. The leading (singular) term is due to the fact that $G^{x(\sing)}$ is independent of $\Delta\vr$, i.e., $G^{x(\sing)}$ enters $g^x$ as the prefactor of the matrix of ones $\vec{v}\vec{v}^\top$.
With this, the left-hand side of Eq.~\eqref{eq:gammaDiag-xp} becomes
\begin{equation}\label{eq:Mperturb}
	M=\sqrt{g^p}\,g^x\sqrt{g^p}=\frac{w}{\lambda}\tilde{\vec{v}}\tilde{\vec{v}}^\top+M^{(\nsing)},
\end{equation}
where $M^{(\nsing)}:=\sqrt{g^p}\,g^{x(\nsing)}\sqrt{g^p}$, $w:=\vec{v}^\top g^p \vec{v}$, and $\tilde{\vec{v}}:=\sqrt{g^p}\,\vec{v}/\sqrt{w}$ such that $\|\tilde{\vec{v}}\|=1$.

In the thermodynamic limit $\lambda\to 0$ ($\Leftrightarrow$ $q_0\to 0$), we can obtain the covariance-matrix eigenvalues $\nu_k$ in Eq.~\eqref{eq:gammaDiag-xp} (the eigenvalues of $M$ are $\nu_k^2$) through perturbation theory in $\lambda$. The unperturbed problem has the eigenvalue $w/\lambda$ for eigenvector $\tilde{\vec{v}}$ and eigenvalue zero in the orthogonal complement of $\tilde{\vec{v}}$. So, degenerate first-order perturbation theory gives the spectrum
\begin{subequations}
\begin{align}
	\nu_\sing^2&=\frac{w}{\lambda}+\tilde{\vec{v}}^\top M^{(\nsing)}\tilde{\vec{v}},\\ \label{eq:nu-nonsing}
	\{\nu_\nsing^2\}&=\operatorname{spect}(PM^{(\nsing)}P),
\end{align}
\end{subequations}
where $P=\id-\tilde{\vec{v}}\tilde{\vec{v}}^\top$ projects onto the orthogonal complement of $\tilde{\vec{v}}$.
\begin{figure*}[t]
	\includegraphics[height=0.3\textwidth]{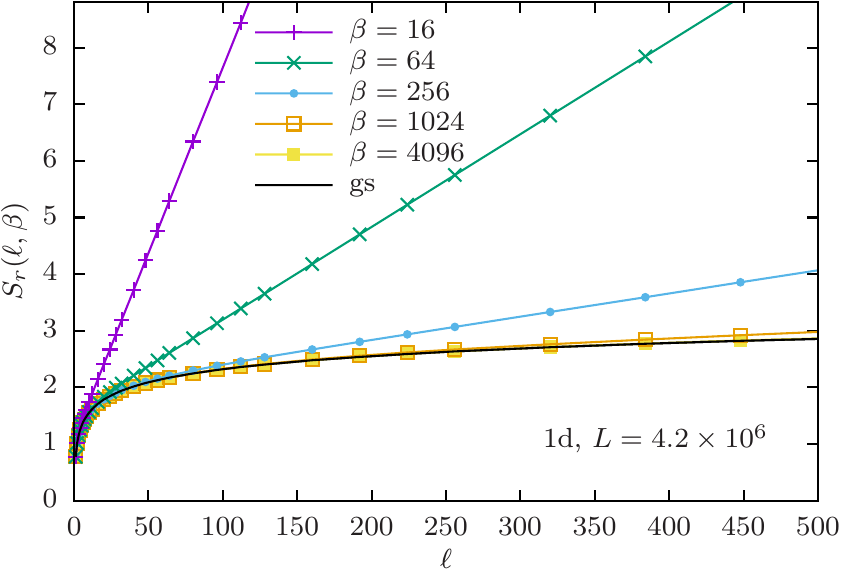}\hspace{8ex}
	\includegraphics[height=0.3\textwidth]{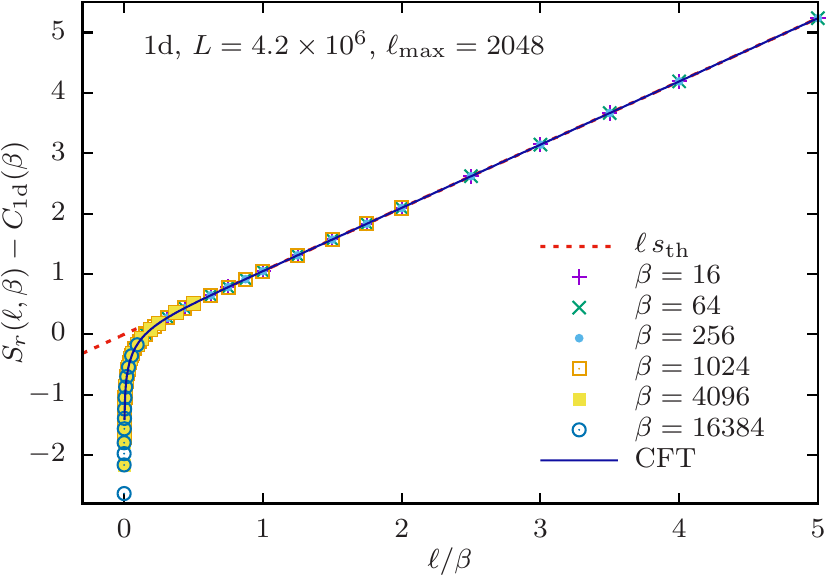}
	\caption{\label{fig:1d}\textbf{Crossover scaling function for 1d.} The regularized subsystem entropies \eqref{eq:Sr1d}, computed for the critical 1d model with various temperatures and subsystem sizes, follow the logarithmic groundstate scaling before crossing over to an extensive regime (left). After subtraction of the subleading term \eqref{eq:ScftC1d}, the data collapse onto a scaling function when plotted as a function of $\ell/\beta$ (right). In this case, the scaling function can be obtained from 1+1d CFT [Eq.~\eqref{eq:ScftScaling}]. In the extensive regime, it is in accordance with the thermodynamic entropy density \eqref{eq:Sth1d}. The total system size was set to $L\approx 4.2\times 10^6$.}
\end{figure*}

For the evaluation of entropies according to Eq.~\eqref{eq:Snu}, the second term in the singular eigenvalue $\nu_\sing^2$ is actually irrelevant: For $y_\sing=\frac{1}{\lambda'}+y+\mc{O}(\lambda')$, we have
\begin{equation*}
	y_\sing\ln y_\sing=\left(\frac{1}{\lambda'}+y\right)\ln\frac{1}{\lambda'}+y+\mc{O}(\lambda')
\end{equation*}
and $h(y_\sing)=\ln\frac{1}{\lambda'}+1+\mc{O}(\lambda')$ is hence $y$ independent. For the subsystem entropy \eqref{eq:Snu}, we finally arrive at the result
\begin{equation}\label{eq:Sfinal}
	S_\A=S_\A^{(\nsing)}+\frac{1}{2}\ln \frac{w}{\lambda}+1,
\end{equation}
where $S_\A^{(\nsing)}$ is the contribution from the nonsingular eigenvalues $\nu_\nsing$ in Eq.~\eqref{eq:nu-nonsing}.

Equation~\eqref{eq:Sfinal}, allows us to extract the entropies for the thermodynamic limit from finite-size computations and to control infrared singularities in 1d and 2d. In particular, we can choose any sufficiently large linear system size $L$ and compute $g^x$ and $g^p$ according to Eq.~\eqref{eq:CovMatrices}. The matrix $g^p$ converges very quickly with $L$. From this, we get converged $w$ and $\tilde{\vec{v}}$ as specified below Eq.~\eqref{eq:Mperturb}. Then, we evaluate $PM^{(\nsing)}P=PMP=P\sqrt{g^p}\,g^{x}\sqrt{g^p}P$, which also converges quickly with $L$, and obtain $\{\nu_\nsing\}$ from it. This yields all terms for Eq.~\eqref{eq:Sfinal} besides $\lambda$. The singular term $1/\lambda\equiv G^{x(\sing)}$, as given by Eq.~\eqref{eq:GxSing}, depends on the infrared cutoff $k_0=q_0/\beta$ but not on the subsystem size.

\section{Entanglement crossovers}\label{sec:Crossovers}
In this section, we determine the scaling functions for the entanglement crossover using finite-temperature subsystem entropies for $\dm=\frac{1}{Z}\,e^{-\beta\hH}$ with $\omega=0$ and $\kappa=1$ as before. As discussed in the introduction, the coincidence of finite-temperature subsystem entropies and the energy-eigenstate entanglement entropies is due to the ETH. The validity of the ETH for the considered systems is substantiated numerically in Sec.~\ref{sec:assertETH}.

\subsection{\texorpdfstring{$d=1$}{d=1}}\label{sec:Crossover1d}
For the critical 1d harmonic lattice model, the singular term \eqref{eq:GxSing1d} in the subsystem entropy has $\lambda=\pi q_0$. Based on Eq.~\eqref{eq:Sfinal}, we can hence define the regularized 1d subsystem entropy
\begin{equation}\label{eq:Sr1d}
	S_r(\ell,\beta)=S^{(\nsing)}(\ell,\beta)+\frac{1}{2}\ln {w}(\ell,\beta)+1,
\end{equation}
where we have subtracted the $q_0$ term.

Figure~\ref{fig:1d} shows the regularized subsystem entropies after subtraction of a subleading $\ell$-independent term
\begin{equation}\label{eq:ScftC1d}
	C_\text{1d}(\beta):=\frac{c}{3}\ln\left(\frac{\beta v}{2\pi a}\right)+c'
\end{equation}
that we discuss below. For several temperatures, the entropies are plotted as functions of $\ell/\beta$ in the right panel. The data collapse onto a single curve -- the scaling function that describes the crossover of subsystem entanglement entropies from the log-area law ($S\sim\ln\ell$) in the ground state to the volume law ($S\sim\ell$) for excited energy eigenstates. The data collapse is due to the scale invariance in this quantum critical regime \cite{Sachdev2011} of the model. The dispersion is linear at low momenta which dominate the long-range physics with group velocity $v=1$. There is just a single energy scale, set by the temperature $\beta^{-1}$. Hence, $S_r-C_\text{1d}$ is not a function of $\ell$ \emph{and} $\beta$ but only of $\ell/\beta$,
\begin{equation}
	S_r(\ell,\beta)-C_\text{1d}(\beta)=\Phi_\text{1d}(\ell/\beta).
\end{equation}

In fact, Poincar\'e and scale invariance in the continuum limit \eqref{eq:H-KG} imply that the long-range physics is described by $(d+1)$-dimensional CFT \cite{Belavin1984-241,Francesco1997,Polchinski1988-383}. For $d=1$, the conformal invariance is very restrictive and the CFT subsystem entropies can be computed using the replica trick and analytic continuation \cite{Korepin2004-92,Calabrese2004}. One obtains
\begin{subequations}
\begin{equation}\label{eq:Scft}
	S^\text{cft} = \frac{c}{3}\ln\left[\frac{\beta v}{\pi a}\sinh\left(\frac{\pi\ell}{\beta v}\right)\right] + c'
\end{equation}
with the central charge being $c=1$ in our case, with an ultraviolet cutoff $1/a$ as set by the lattice spacing, and a nonuniversal constant $c'$. The universal scaling function $\Phi_\text{1d}$ is simply the leading term in
\begin{equation}\label{eq:ScftScaling}
	S^\text{cft} = \frac{c}{3}\ln\left[2\sinh \left(\frac{\pi}{v}\,\frac{\ell}{\beta}\right)\right]+\mc{O}(\ell^0),
\end{equation}
\end{subequations}
which is indeed a function of $\ell/\beta$ only. And we can read off the subleading term \eqref{eq:ScftC1d},
which was taken into account for Fig.~\ref{fig:1d}.
\begin{figure*}[t]
	\includegraphics[height=0.3\textwidth]{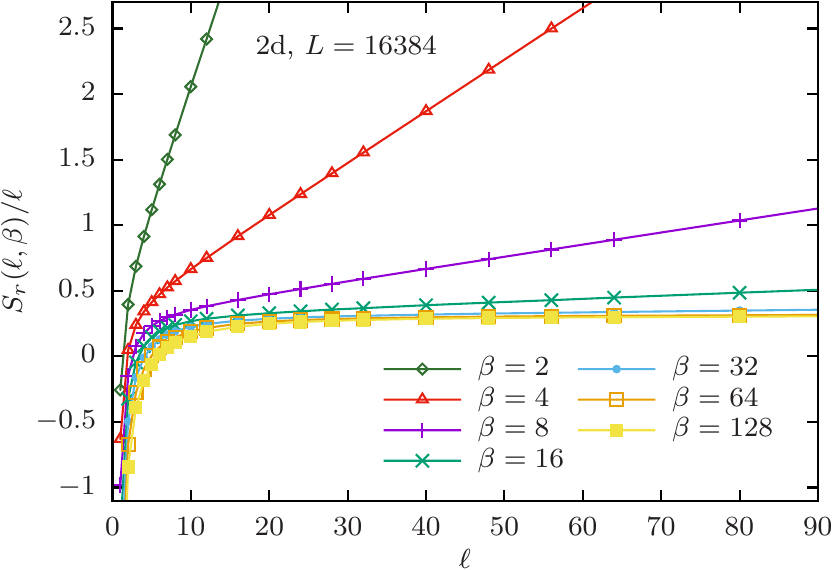}\hspace{8ex}
	\includegraphics[height=0.3\textwidth]{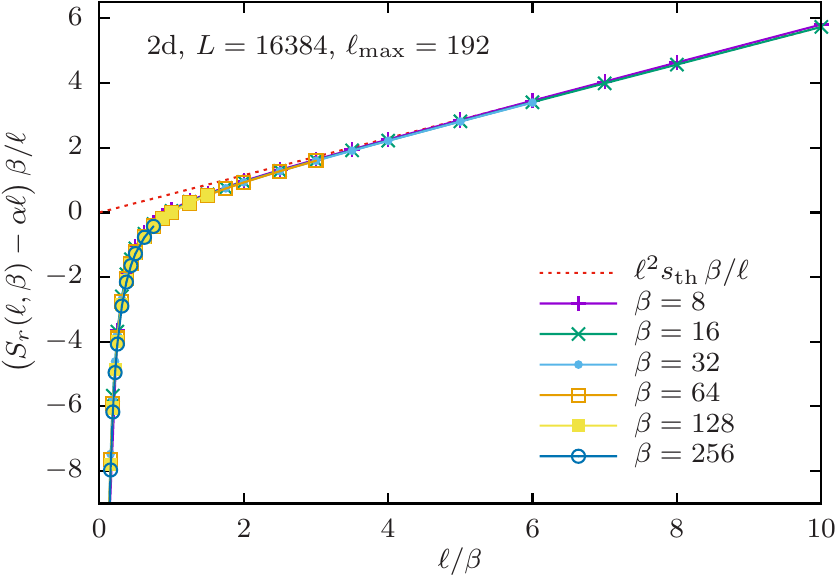}
	\caption{\label{fig:2d}\textbf{Crossover scaling function for 2d.} This figure shows regularized subsystem entropies \eqref{eq:Sr2d} for the critical 2d model, computed for various temperatures and subsystems of size $\ell\times\ell$ (left). When plotted as a function of $\ell/\beta$, the data collapse onto a scaling function (right). The total system has $L\times L$ sites with $L=\num{16384}$. In the extensive large-$\ell/\beta$ regime, the subsystem entropy is in accordance with the thermodynamic entropy density \eqref{eq:Sth2d}.}
\end{figure*}

For small subsystem size $\ell$ or temperature $\beta^{-1}$, Eq.~\eqref{eq:Scft} recovers the log law $\frac{c}{3}\ln(\ell/a)$ of the groundstate entanglement entropy. This can be attributed to a limited momentum-space resolution when probing on small subsystems \cite{Miao2019_05}. The crossover to extensive scaling $S\sim \ell$ occurs at $\ell\sim\beta v/\pi$. The corresponding entropy density can be deduced by evaluation of the thermodynamic entropy $S_\therm$. In Appendix~\ref{sec:Sth}, we show this computation and the 1d result
\begin{equation}\label{eq:Sth1d}
	s_\therm=\frac{S_\therm}{L}=\frac{1}{\beta}\,\frac{\pi}{3}
\end{equation}
indeed coincides with the large-$\ell/\beta$ limit of Eq.~\eqref{eq:ScftScaling}.

\subsection{\texorpdfstring{$d=2$}{d=2}}\label{sec:Crossover2d}
For the critical 2d harmonic lattice model, the singular term \eqref{eq:GxSing2d} in the subsystem entropy has $\lambda=2\pi\beta/\ln(1/q_0)$. Based on Eq.~\eqref{eq:Sfinal}, we can hence define the regularized 2d subsystem entropy
\begin{equation}\label{eq:Sr2d}
	S_r(\ell,\beta)= S^{(\nsing)}(\ell,\beta) +\frac{1}{2}\ln {w}(\ell,\beta)+1-\frac{1}{2}\ln(2\pi\beta),
\end{equation}
where we have subtracted the $q_0$ term.

Figure~\ref{fig:2d}, shows regularized subsystem entropies after subtraction of a subleading area-law term $\alpha\ell$, where we extracted $\alpha\approx 0.4464$ from the groundstate entanglement entropy. Subsystems were chosen as $\ell\times\ell$ squares. As in the 1d case, when plotted over $\ell/\beta$, the data for various $\ell$ and $\beta$ collapse onto a scaling function that describes the crossover of subsystem entanglement entropies from the area law ($S\sim\ell$) in the ground state to the volume law ($S\sim\ell^2$) for excited eigenstates. Due to the scale invariance in the quantum critical regime, $S_r-\alpha\ell$ is not a function of $\ell$ \emph{and} $\beta$ but only of $\ell/\beta$,
\begin{equation}
	S_r(\ell,\beta)-\alpha\ell=\Phi_\text{2d}(\ell/\beta).
\end{equation}
In the figure, we multiply it by $\beta/\ell$ to nicely show the crossover to the extensive scaling.

For 2d, the divergent term $\frac{1}{2}\ln(1/\lambda)$ in the subsystem entropy \eqref{eq:Sfinal} grows very slowly (double-logarithmically) in the system size $L$, because $\lambda$ already decreases logarithmically. Hence, using the perturbative approach explained in Sec.~\ref{sec:Singular} is imperative in this case. One could not realistically reach the thermodynamic regime and extract the scaling function with a naive numerical computation.

For small $\ell/\beta$, the subsystem entropies follow the groundstate entanglement curve. For large $\ell/\beta$ we cross over to extensive scaling $S\sim \ell^2$. The corresponding entropy density coincides with the thermodynamic value. As shown in Appendix~\ref{sec:Sth}, it is
\begin{equation}\label{eq:Sth2d}
	s_\therm=\frac{S_\therm}{L^2}=\frac{1}{\beta^2}\frac{3\zeta(3)}{2\pi}
\end{equation}
with the Riemann zeta function $\zeta(s)$.

\section{Squeezed-state excitations}\label{sec:Squeezed}
\begin{figure*}[t]
	\includegraphics[width=0.43\textwidth]{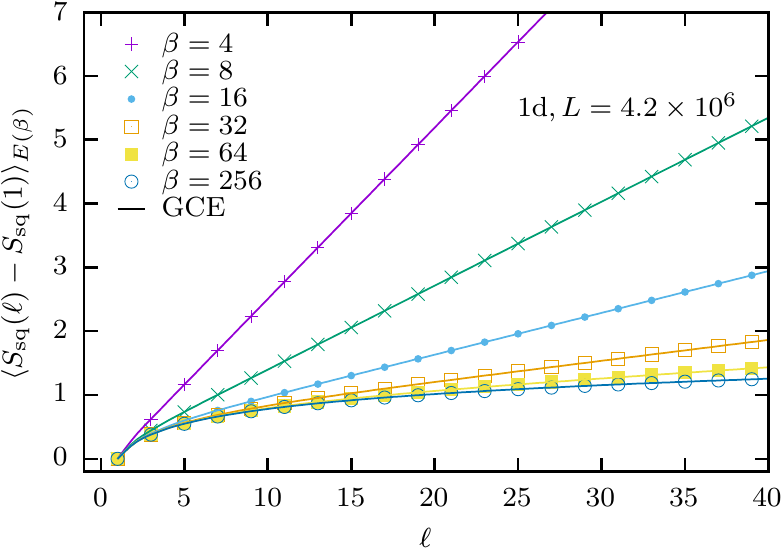}\hspace{8ex}
	\includegraphics[width=0.43\textwidth]{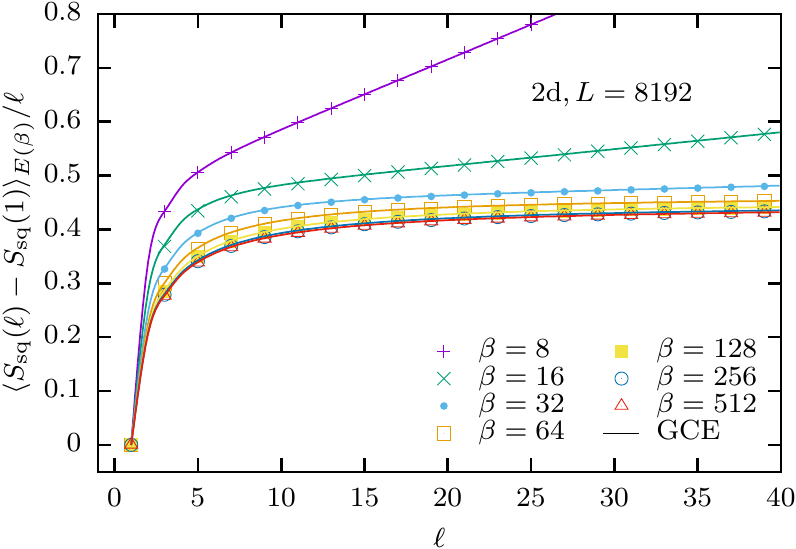}
	\caption{\label{fig:confirmETH}\textbf{Checking ETH for 1d and 2d.} The numerical tests support the applicability of ETH for the study of eigenstate entanglement entropies in the critical harmonic lattice models for 1d (left) and 2d (right). Points are averaged entanglement entropies $S_\sq(\ell)$ for the randomly sampled squeezed states \eqref{eq:squeezedMB} in energy windows of size $\Delta E=1$ around $E(\beta)=\bra\hH\ket_\beta$ and with particle number $\sum_\vk n_\vk=\lfloor\bra\hN\ket_\beta\rfloor$. Standard deviations are smaller than the symbol sizes. Lines show the subsystem entropy $S(\ell,\beta)$ in the corresponding GCE with chemical potential $\mu=0$. The agreement is excellent. For better discriminability and to cancel infrared singularities, we plot $S_\sq(\ell)-S_\sq(1)$. Total linear system sizes are $L=4.2\times 10^6$ for 1d and $L=\num{8192}$ for 2d.}
\end{figure*}
To establish the ETH for the harmonic lattice model \eqref{eq:HLM-r}, we should, in principle, compute entanglement entropies for energy eigenstates
\begin{equation}\label{eq:Eigenstates}
	|\vn\ket=\prod_\vk \frac{1}{\sqrt{n_\vk!}}\big(\hb^\dag_\vk\big)^{n_\vk}|0\ket
\end{equation}
of a fixed energy $E=E(\beta)$ and show that they are sharply peaked around the subsystem entropy $S(\ell,\beta)$ of the corresponding thermal ensemble $\dm=\frac{1}{Z}\,e^{-\beta\hH}$ discussed in Sec.~\ref{sec:Crossovers}. In contrast to quasi-free fermionic systems, the bosonic Fock states \eqref{eq:Eigenstates} are, however, \emph{not} Gaussian states and there are no efficient methods to compute entanglement entropies of large subsystems for these states. The cost would scale exponentially in the subsystem size.

We resolve this problem by studying many-body squeezed states. As discussed below, they are useful approximations to the exact eigenstates \eqref{eq:Eigenstates} with a narrow energy distribution, and they are Gaussian states such that entanglement entropies can be computed efficiently with the method of Sec.~\ref{sec:Entropy}.

\subsection{Diagonalization of \texorpdfstring{$\hH$}{H} with a real transformation}\label{sec:DiagReal}
We will put each $\vk$ mode into a squeezed state. If we would simply squeeze the vacuum state $|0\ket$ with operators $\exp\big(\frac{1}{2}[z^*\hb_\vk^2-z\,(\hb_\vk^\dag)^2]\big)$, the resulting state would actually not be translation invariant, since the corresponding position and momentum operators from Eq.~\eqref{eq:xk-pk} are not self-adjoint, $\hx_\vk^\dag\neq\hx_\vk$ and $\hp_\vk^\dag\neq\hp_\vk$.

To construct translation-invariant squeezed states, we slightly modify the diagonalization procedure, defining new self-adjoint operators through a rotation
\begin{gather*}
	\tvec{\hx'_\vk\\\hx'_{-\vk}}:=U\tvec{\hx_\vk\\\hx_{-\vk}},\quad
	\tvec{\hp'_\vk\\\hp'_{-\vk}}:=U\tvec{\hp_\vk\\\hp_{-\vk}}\quad\\
	\text{with}\quad U=\frac{1}{i\sqrt{2}}\tmat{e^{-i\pi/4}&& e^{i\pi/4}\\ e^{i\pi/4}&& -e^{-i\pi/4}}.
\end{gather*}
These obey the canonical commutation relations \eqref{eq:CCR-k} and can also be written in the form
\begin{align*}
	\hx'_\vk &= \sqrt{\frac{2}{\N}}\sum_\vr \sin\left(\vk\cdot\vr+\pi/4\right)\hx_\vr= \hx_{\vk}^{\prime\dag},\\
	\hp'_\vk &= \sqrt{\frac{2}{\N}}\sum_\vr \sin\left(\vk\cdot\vr+\pi/4\right)\hp_\vr= \hp_{\vk}^{\prime\dag}
\end{align*}
Due to the symmetry $\veps_\vk=\veps_{-\vk}$, we still have $\hH=\frac{1}{2}\sum_{\vk}(\hp^{\prime 2}_\vk+\veps_\vk\hx^{\prime 2}_\vk)$ as in Eq.~\eqref{eq:HLM-Hk}. Finally, defining ladder operators
\begin{equation*}
	\ha_\vk:=\frac{1}{\sqrt{2}}\left(\veps_\vk^{1/2}\hx'_\vk+i\veps_\vk^{-1/2}\hp'_\vk\right),
\end{equation*}
we arrive at $\hH=\sum_{\vk}\veps_\vk\big(\ha_\vk^\dag\ha^\pdag_\vk+1/2\big)$ as in Eq.~\eqref{eq:HLM-diag}. Clearly, the vacuum $|0\ket$ for the annihilation operators $\hb_\vk$ is also the vacuum for the $\ha_\vk$.

\subsection{Squeezed states: Single mode}
For a single bosonic mode with ladder operator $\ha$, we define the squeeze operators $\hS(z)$ and squeezed states $|z\ket$,
\begin{equation}\label{eq:squeezed1}
	\hS(z):=e^{\frac{1}{2}\left[z^*\ha^2-z\,(\ha^\dag)^2\right]},\,\,\,\,\,
	|z\ket:=\hS(z)\,|0\ket\,\,\,\,\, \forall z\in\mathbb{C}
\end{equation}
with unitary $\hS^\dag(z)=\hS(-z)$.
With the polar form $z=r e^{i\vphi}$, let us also introduce the squeezed operators
\begin{equation*}
	\ha(z):=\hS^\dag(z)\,\ha\,\hS(z)
	=\cosh(r)\,\ha-e^{i\vphi}\sinh(r)\,\ha^\dag,
\end{equation*}
which obey the canonical commutation relations.

For the computation of entanglement entropies we will need expectation values $\bra\ha^\dag\ha\ket_z$ and $\bra\ha\ha\ket_z$. The first is
\begin{subequations}\label{eq:squeeze1G}
\begin{equation}
	\bra\ha^\dag\ha\ket_z = \bra\ha^\dag(z)\ha(z)\ket_0
	=\sinh^2(r),
\end{equation}
where $\bra\dots\ket_0$ denotes vacuum expectation values. To mimic particle number eigenstates $|n\ket$ as in Eq.~\eqref{eq:Eigenstates}, we will choose integer integer occupation-number expectation values
\begin{equation}\nonumber
	\bra\ha^\dag\ha\ket_z=n\in\mathbb{N}\quad\Leftrightarrow\quad
	r=\operatorname{asinh}\big(\sqrt{n}\big).
\end{equation}
The second required expectation value is
\begin{align}\nonumber
	\bra\ha\ha\ket_z &= \bra\ha(z)\ha(z)\ket_0
	=-e^{i\vphi}\sinh(r)\cosh(r)\\
	&=-e^{i\vphi}\sqrt{n\,(n+1)}.
\end{align}
\end{subequations}
With Wick's theorem, one finds the particle number standard deviation to be $\Delta n_z=\sqrt{\bra\hat{n}^2\ket_z-n^2}=\sqrt{2n(n+1)}$.

\subsection{Squeezed states: Many-body covariances}\label{sec:SqueezedMB}
In generalization of Eq.~\eqref{eq:squeezed1}, we employ many-body squeezed states
\begin{subequations}\label{eq:squeezedMB}
\begin{equation}
	|\vz\ket:=\prod_\vk\hS(z_\vk)\,|0\ket \quad\text{with}\quad z_\vk=r_\vk e^{i\vphi_\vk}\in\mathbb{C}.
\end{equation}
and choose
\begin{equation}
 	z_\vk=z_{-\vk} \quad\text{and}\quad
 	r_\vk=\operatorname{asinh}\big(\sqrt{n_\vk}\big).
\end{equation}
\end{subequations}
They are Gaussian states and good approximations to the corresponding Fock states $|\vn\ket$ in Eq.~\eqref{eq:Eigenstates}. In particular, they have translation-invariant covariance matrices, agreeing occupation number expectation values $\bra\ha_\vk^\dag\ha_\vk^\pdag\ket_\vz=n_\vk$, and vanishing first moments $\bra\ha_\vk\ket_\vz=0$.
Hence, also the total particle number and energy expectation values coincide, $\bra\hN\ket_\vz=\sum_\vk n_\vk$ and $\bra\hH\ket_\vz=\sum_\vk \veps_\vk n_\vk$. The relative fluctuations of these quantities vanish in the thermodynamic limit. As an example, we show in Appendix~\ref{sec:SqueezedDeltaN} that $\Delta N_\vz/N_\vz$ scales in the typical states as $1/\ln L$ for 1d, $\sqrt{\ln L}/L$ for 2d, and $L^{-d/2}$ for $d\geq 3$.

The covariance matrices $G^x_{\vr,\vr'}=\bra\hx_\vr\hx_{\vr'}\ket_\vz$, $G^p_{\vr,\vr'}=\bra\hp_\vr\hp_{\vr'}\ket_\vz$ and $G^{xp}_{\vr,\vr'}=\bra\hx_\vr\hp_{\vr'}\ket_\vz$, needed for the computation of entanglement entropies according to Sec.~\ref{sec:Entropy}, follow from Eqs.~\eqref{eq:squeeze1G}. With $\Delta\vr=\vr-\vr'$, one obtains
\begin{multline*}\textstyle
	G^x_{\vr,\vr'}
	 = \frac{1}{\N}\sum_{\vk} \frac{1}{\veps_\vk}\,\cos(\vk\cdot\Delta\vr)\\\textstyle
	 \times\left(n_\vk+\frac{1}{2}-\cos(\vphi_\vk)\, \sqrt{n_\vk\,(n_\vk+1)}\right),
\end{multline*}
\begin{multline*}\textstyle
	G^p_{\vr,\vr'}
	 = \frac{1}{\N}\sum_{\vk} \veps_\vk\,\cos(\vk\cdot\Delta\vr)\\\textstyle
	 \times\left(n_\vk+\frac{1}{2}+\cos(\vphi_\vk)\, \sqrt{n_\vk\,(n_\vk+1)}\right),
\end{multline*}
\begin{equation*}\textstyle
	G^{xp}_{\vr,\vr'}
	 = \frac{1}{\N}\sum_{\vk} \cos(\vk\cdot\Delta\vr)
	 \left(\frac{i}{2}-\sin(\vphi_\vk)\, \sqrt{n_\vk\,(n_\vk+1})\right).
\end{equation*}
\begin{figure}[t]
	\includegraphics[width=\columnwidth]{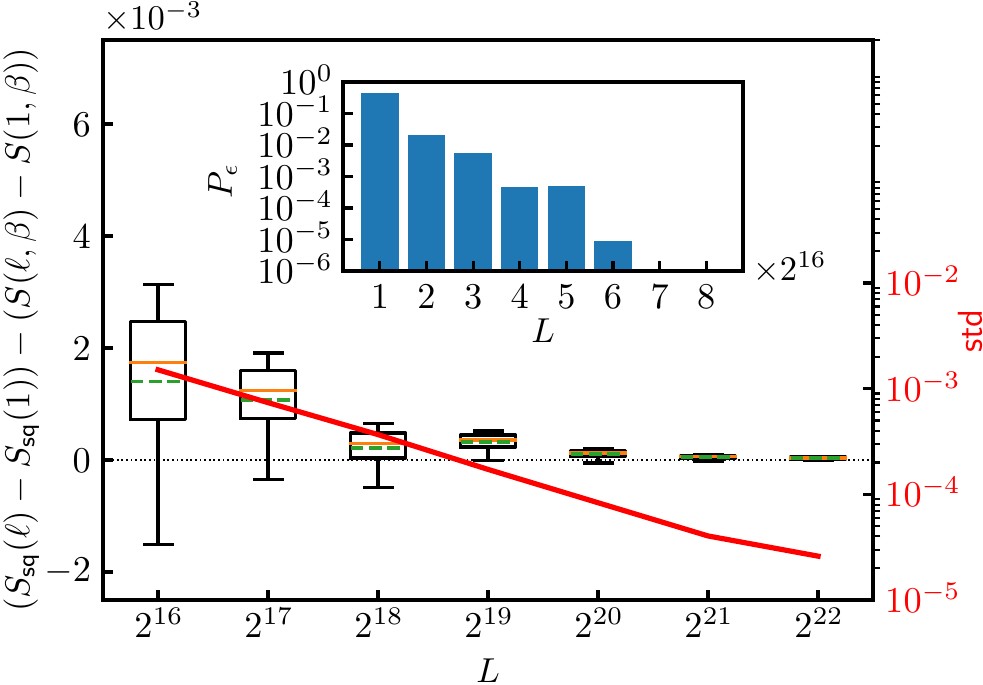}
	\caption{\label{fig:EEdistribution}\textbf{ETH and entanglement distribution.} For the critical 1d harmonic lattice model, this plot characterizes the distribution of entanglement entropies $S_\sq(\ell)$ for randomly sampled squeezed states \eqref{eq:squeezedMB} in energy windows of size $\Delta E=1$ around $E(\beta)=\bra\hH\ket_\beta$ with $\beta=16$ and particle number $\sum_\vk n_\vk=\lfloor\bra\hN\ket_\beta\rfloor$. For subsystem size $\ell=32$ and for each of the considered total sizes $L=2^{16},\dotsc,2^{22}$, about $\num{115000}$ samples were generated, each corresponding to $16\N$ updates. The main panel shows the 5\%, 25\%, 50\%, 75\%, and 95\% quantiles (lower whisker, box bottom, orange line, box top, upper whisker) of the distribution, where the GCE value has been subtracted. Dashed green lines show the averages. The standard deviation (red) is plotted double-logarithmically. The inset shows the ratio \eqref{eq:Poutlier} of untypical states for a deviation threshold of $\epsilon=0.002$.}
\end{figure}

\section{Assertion of the ETH in one and two dimensions}\label{sec:assertETH}
To assert the ETH, we sample squeezed states \eqref{eq:squeezedMB} with integer $n_\vk$ obeying $\sum_\vk n_\vk=\lfloor\bra\hN\ket_\beta\rfloor$ from small energy windows of width $\Delta E$ around $E(\beta):=\bra\hH\ket_\beta$ with equal probability. For each sample, angles $\vphi_\vk\in[0,2\pi)$ are chosen randomly according to the uniform distribution. A Markov chain of such squeezed states is generated as follows. We start from an arbitrary valid initial choice for the $\{n_\vk\}$. In each update, two wavevectors $\vk$ and $\vk'$ are chosen randomly. For $n_\vk \geq 1$, we decrease $n_\vk$ and increase $n_{\vk'}$ by 1 if the energy stays in the predefined window. Otherwise, the update is rejected. For each energy $E(\beta)$, a total of $\num{16000}\N$ updates are done, and entanglement entropies are computed every $16\N$ iterations, where $\N=L^d$ is the total system size.

ETH can be explained through a momentum-space resolution limitation: With observations on a subsystem of linear size $\ell$, one cannot resolve variations of
momentum-space Green's functions below a scale $\sim 1/\ell$. Typical samples have $n_\vk$ very close to the GCE expectation value $\bra\hat{n}_\vk\ket_\beta$, in the sense that $n_\vk$, averaged over small momentum-space patches, quickly approaches $\bra\hat{n}_\vk\ket_\beta$. These are then smooth functions of $\vk$. See the appendix of Ref.~\cite{Miao2019_05} for details. This argument applies to the actual eigenstates \eqref{eq:Eigenstates} as well as their squeezed-state approximations \eqref{eq:squeezedMB} that we employ.
For the latter, also note that the averages $\overline{\cos\vphi_\vk}$ and $\overline{\sin\vphi_\vk}$ over small momentum-space patches vanish for random angles $\vphi_\vk$, such that the distribution of the squeezed-state covariance matrices $G$ in Sec.~\ref{sec:SqueezedMB} will be sharply peaked around the GCE covariance matrices \eqref{eq:CovMatrices}. Hence, the entanglement entropies $S_\sq(\ell)$ of these squeezed-state excitations will be sharply peaked around the GCE subsystem entropies $S(\ell,\beta)$ analyzed in Sec.~\ref{sec:Crossovers}.

For the quantum-critical 1d and 2d systems, we again have the infrared singularity in $G^x$ as discussed in Sec.~\ref{sec:Singular}. It results here in a sensitivity with respect to the angles $\vphi_\vk$ for small $|\vk|$. This effect can, however, be removed through a regularization. Since we are not interested in scale invariance here, we simply regularize by considering the differences $S_\sq(\ell)-S_\sq(1)$.

The above arguments on the coincidence of $S_\sq(\ell)$ and $S(\ell,\beta)$ are confirmed numerically in Fig.~\ref{fig:confirmETH} for the 1d and 2d critical harmonic lattice models [$\omega=0$ and $\kappa=1$ in Eq.~\eqref{eq:HLM-r}], showing excellent agreement. In these plots, the standard deviations for $S_\sq(\ell)$ are smaller than the symbol sizes.
An analysis of the $S_\sq(\ell)$ distribution is given in Fig.~\ref{fig:EEdistribution} for the critical 1d model and a fixed subsystem size $\ell=32$. The main panel shows various quantiles of the $S_\sq(\ell)$ distribution. With increasing system size $L$, the average converges to the GCE value, and the distribution becomes more and more narrow. In particular, a double-logarithmic plot of the standard deviation displays a $1/L$ decrease. The inset provides a large-deviation analysis \cite{Touchette2009-478,Yoshizawa2018-120}, showing the ratio of untypical states, defined as 
\begin{equation}\label{eq:Poutlier}
	P_\epsilon=P(\big|[S_\sq(\ell)-S_\sq(1)]-[S(\ell,\beta)-S(1,\beta)]\big|>\epsilon).
\end{equation}
The semi-logarithmic plot for threshold $\epsilon=0.002$ indicates an exponential decay of $P_\epsilon$ in $L$. Deviations at large $L$ (visible for $L\gtrsim 5\times 2^{16}$) can be attributed to the limited number of samples. Such an exponential decay is indeed expected for the weak ETH in integrable systems, whereas strong ETH for non-integrable systems should result in a double-exponential decay \cite{Yoshizawa2018-120}.

\section{Conclusion}\label{sec:Conclusion}
In conclusion, the ETH can be applied for harmonic lattice models to elucidate the crossover of entanglement entropies in energy eigenstates from the groundstate scaling at small subsystem sizes and low energies to the extensive scaling at large sizes and higher energies. In particular, entanglement entropies of almost all eigenstates coincide with subsystem entropies of a corresponding thermal equilibrium ensemble. A large-deviation analysis for the critical 1d model shows that the ratio of athermal eigenstates decays exponentially with increasing system size.
We find scaling functions for the crossovers in critical 1d and 2d harmonic lattices (massless free scalar quantum field theory). To this purpose, we also introduced an infrared regularization scheme that retains scale invariance. The obtained scaling functions should also apply to the quantum critical regions of interacting systems whose renormalization-group fixed point is the free scalar theory \cite{Sachdev2011}. It would be very valuable to derive analytical expressions for the crossover functions.

The results, shown here for von Neumann entanglement entropies \eqref{eq:vNentang}, do also apply for R\'{e}nyi entanglement entropies. The latter can, e.g., be used to deduce upper bounds on computation costs of tensor network simulations \cite{Verstraete2005-5,Barthel2017_08}.

We gratefully acknowledge discussions with Pasquale Calabrese and Marcos Rigol, as well as support through US Department of Energy grant DE-SC0019449.

\appendix

\section{Thermodynamic entropy densities}\label{sec:Sth}
Eigenstate entanglement entropies are related to thermodynamic subsystem entropies due to ETH as discussed in the introduction. As a function of the linear subsystem size $\ell$, there is a temperature-dependent crossover to a volume law $S\sim \ell^d s_\therm(\beta)$. In the following, we determine the thermodynamic entropy densities $s_\therm$ for the critical harmonic lattice model \eqref{eq:HLM-diag}. In particular, we consider the low-energy regime as described by the free scalar field theory \eqref{eq:H-KG} with the linear dispersion relation $\veps_\vk=|\vk|$.

For the grand-canonical ensemble $\dm=\frac{1}{Z}\,e^{-\beta\hH}$, one finds the following well-known result for the thermodynamic entropy
\begin{equation}\label{eq:Sth_general}
\begin{aligned}
	S_\therm
	&=-\Tr\dm\ln\dm\\
	&=\sum_\vk\big[(\tn_\vk+1)\ln(\tn_\vk+1)-\tn_\vk\ln \tn_\vk\big]
\end{aligned}
\end{equation}
with the Bose-Einstein distribution
\begin{equation}\label{eq:BED}
	\tn_\vk:=\bra\hat{n}_\vk\ket_\beta=1/\left(e^{\beta\veps_\vk}-1\right).
\end{equation}

The density of states for the linear dispersion relation $\veps_\vk=|\vk|$ is
\begin{subequations}\label{eq:DOS}
\begin{equation}
	g(\veps)=\frac{1}{(2\pi)^d}\int\ud^d k\,\delta(\veps_\vk-\veps)
	=
	\begin{cases}
	 \frac{1}{\pi}		&\text{for 1d,}\\
	 \frac{\veps}{2\pi}	&\text{for 2d,}\\
	 \frac{\veps^2}{2\pi^2}	&\text{for 3d,}
	\end{cases}
\end{equation}
and $\veps\geq 0$. With the Heaviside step function $\Theta(\veps)$, we can write it as
\begin{equation}
	g(\veps)=:g_d \,\veps^{d-1}\Theta(\veps).
\end{equation}
\end{subequations}

Taking the thermodynamic limit, $\sum_\vk\mapsto L^d\int\ud\veps\, g(\veps)$ in Eq.~\eqref{eq:Sth_general}, and substituting $q:=\beta\veps$, we obtain
\begin{equation*}
	S_\therm=\frac{g_d L^d}{\beta^d}\int_0^\infty\ud q\, q^{d-1}\big[(\tn_q+1)\ln(\tn_q+1)-\tn_q\ln \tn_q\big]
\end{equation*}
with $\tn_q=1/\left(e^{q}-1\right)$. The integrals can be done analytically, giving in the thermodynamic entropy densities
\begin{equation}
	s_\therm(\beta)=\frac{S_\therm}{L^d}=
	\begin{cases}
	\begin{aligned}
	 &\frac{1}{\beta}\frac{\pi}{3}		 &\!\!\!\!\approx\frac{1.047}{\beta}	&\quad\text{for 1d,}\\
	 &\frac{1}{\beta^2}\frac{3\zeta(3)}{2\pi}&\!\!\!\!\approx\frac{0.574}{\beta^2}	&\quad\text{for 2d,}\\
	 &\frac{1}{\beta^3}\frac{2\pi^2}{45}	 &\!\!\!\!\approx\frac{0.439}{\beta^3}	&\quad\text{for 3d.}
	\end{aligned}
	\end{cases}
\end{equation}

\section{Squeezed-state particle-number fluctuations}\label{sec:SqueezedDeltaN}
To assess the validity of the ETH for entanglement entropies in the critical harmonic lattice models, we employed squeezed states $|\vz\ket$ [Eqs.~\eqref{eq:squeezedMB}] to approximate the exact eigenstates $|\vn\ket$ [Eq.~\eqref{eq:Eigenstates}]. As discussed in Sec.~\ref{sec:SqueezedMB}, they are good approximations in the sense that they have translation-invariant covariance matrices, vanishing first moments, and agreeing occupation number expectation values. Hence, also their expectation values for the total particle number operator and Hamiltonian agree with those of the corresponding Fock states $|\vn\ket$. Relative fluctuations of these observables vanish in the thermodynamic limit. This is exemplified here by evaluating the relative particle number fluctuations $\Delta N_\vz/N_\vz$ in the typical states which, after coarse-graining in momentum space, have occupation numbers according to the Bose-Einstein distribution \eqref{eq:BED}, i.e., $\vn\to\tilde{\vn}$.

Using Wick's theorem \cite{Fetter1971} and Eqs.\eqref{eq:squeeze1G}, the particle number variance $\Delta N_\vz^2\equiv\bra\hN^2\ket_\vz -\bra\hN\ket_\vz^2$ can be written in the form
\begin{align}\nonumber
	\Delta N_\vz^2 &=\sum_{\vk,\vk'}\left(\bra\hat{n}_\vk\hat{n}_{\vk'}\ket_\vz-n_\vk n_{\vk'}\right)
	=\sum_{\vk}2n_\vk(n_\vk+1)\\
	& \xlongrightarrow{\text{coarse-grain}}
	   \sum_{\vk}2\tn_\vk(\tn_\vk+1).
\end{align}
In the limit of large system size $\N=L^d$, with the density of states $g(\veps)$ as in Eq.~\eqref{eq:DOS}, $q:=\beta\veps$ and an infrared cutoff $q_0=:2\pi \beta/L$ as in Sec.~\ref{sec:Singular}, we have
\begin{align}\label{eq:sum_nk}
	\sum_\vk \tn_\vk   &\xlongrightarrow{\text{large}\ L}
	 \frac{g_d L^{d}}{\beta^d}\int_{q_0}^\infty\ud q\, \frac{q^{d-1}}{e^q-1},\\ \label{eq:sum_nk2}
	\sum_\vk \tn_\vk^2 &\xlongrightarrow{\text{large}\ L}
	 \frac{g_d L^{d}}{\beta^d}\int_{q_0}^\infty\ud q\, \frac{q^{d-1}}{\left(e^q-1\right)^2}.
\end{align}
The integral in Eq.~\eqref{eq:sum_nk} is convergent for $d\geq 2$ dimensions, giving $\approx 1.645$ for $d=2$ dimensions and $\approx 2.404$ for $d=3$. The infrared divergent contribution for $d=1$ is $\sim -\ln q_0\sim\ln L$. The integral in Eq.~\eqref{eq:sum_nk2} is convergent for $d\geq 3$, giving $\approx 0.885$ for $d=3$ dimensions. The infrared divergent contribution for $d=1$ is to leading order $\sim 1/q_0= L/(2\pi\beta)$ and, for $d=2$, it is $\sim \ln L$.

For the $L$ dependence of the relative particle number fluctuations we hence have
\begin{equation*}
	\frac{\Delta N_\vz}{N_\vz}=\frac{\sqrt{\sum_\vk 2n_\vk(n_\vk+1) }}{\sum_\vk n_\vk} \sim
	\begin{cases}
	 1/\ln L		&\text{for}\,d=1,\\
	 \sqrt{\ln L}/L		&\text{for}\,d=2,\\
	 L^{-d/2}		&\text{for}\,d\geq 3.
	\end{cases}
\end{equation*}
So, for any number of dimensions, the relative fluctuations vanish in the thermodynamic limit.

\end{document}